\documentclass[american,aps,pre,showpacs,twocolumn]{revtex4}
\usepackage{graphicx}
\usepackage{amsmath}
\usepackage{color}
\makeatletter
\usepackage{babel}
\makeatother
\begin{document}

\title{Percolation phase transition by removal of $k^{2}$-mers from fully occupied lattices}

\author{L.S. Ramirez$^{*}$, P.M. Centres, A.J. Ramirez-Pastor}
\affiliation{Departamento de F\'{\i}sica, Instituto de F\'{\i}sica Aplicada, Universidad Nacional de San Luis-CONICET, Ej\'ercito de Los Andes 950, D5700HHW, San  Luis, Argentina}


\email{lsramirez@unsl.edu.ar}

\begin{abstract}
Numerical simulations and finite-size scaling analysis have been carried out to study the problem of inverse site percolation by the removal of $k \times k$ square tiles ($k^{2}$-mers) from square lattices. The process starts with an initial configuration, where all lattice sites are occupied and, obviously, the opposite sides of the lattice are connected by occupied sites. Then, the system is diluted by removing $k^{2}$-mers of occupied sites from the lattice following a random sequential adsorption mechanism. The process finishes when the jamming state is reached and no more objects can be removed due to the absence of occupied sites clusters of appropriate size and shape. The central idea of this paper is based on finding the maximum concentration of occupied sites, $p_{c,k}$, for which the connectivity disappears. This particular value of the concentration is called \textit{inverse percolation threshold}, and determines a well-defined geometrical phase transition in the system. The results obtained for $p_{c,k}$ show that the inverse percolation threshold is a decreasing function of $k$ in the range $1 \leq k \leq 4$. For $k \geq 5$, all jammed configurations are percolating states, and consequently, there is no non-percolating phase. In other words, the lattice remains connected even when the highest allowed concentration of removed sites is reached. 
The jamming exponent $\nu_j$ was measured, being $\nu_j = 1$ regardless of the size $k$ considered. In addition, the accurate determination of the critical exponents $\nu$, $\beta$ and $\gamma$ reveals that the percolation phase transition involved in the system, which occurs for $k$ varying between 1 and 4, has the same universality class as the standard percolation problem.

\end{abstract}

\pacs{82.35.Np, 68.47.Pe, 82.35.Gh,81.15.-z, 61.30.-v}

\maketitle


Keywords: jamming, percolation, RSA, phase transitions\\

\section{Introduction} \label{introduccion}

The percolation theory is one of the simplest models in probability theory, which has been applied to a wide range of phenomena in physics, chemistry, biology, and materials science where connectivity and clustering play an important role: flow in porous materials \cite{Stauffer,Sahimi,Li2014}, network theory \cite{Dorogo,Newman,Newman1,Cohen,Kornbluth,Lowinger}, thermal phase transitions \cite{Gao,Coniglio}, spread of the computer virus \cite{Kenah}, transport in disordered media \cite{Yazdi,Kirkpatrick}, electrical conductivity in alloys \cite{Chatter,Tara2016,Lebo2018,Tara2018a}, simulated spread fire in multi-compartmented structures \cite{Zekri} and the spread of epidemics \cite{Miller}. Percolation theory has also provided insight into the behavior of more complicated models exhibiting phase transitions and critical phenomena \cite{Stauffer,Sahimi,Grimmett,Christensen,Bollobas}.

Classical percolation theory studies site and bond percolation. In the case of discrete lattices, each cell is a site and the bond is edge between cells. Then, a single position (site/bond) is occupied with probability $p$. For a precise value of $p$, a cluster of nearest-neighbor sites (bonds) extends from one side to the opposite side of the system. This particular value of concentration rate is named percolation threshold $p_c$. At this critical concentration a second-order phase transition occurs, which is characterized by well-defined critical exponents \cite{Stauffer}.

Percolation theory can also be used to understand network robustness, i.e., how the structure of a network changes as its elements (sites/bonds) are removed either through random or malicious attacks \cite{Dorogo,Newman,Newman1,Cohen,Kornbluth,Lowinger}. The focus of robustness in complex networks is the response of the network to the removal of nodes or links. The model of such a process can be thought of as an inverse percolation problem. The term inverse is used simply to indicate that the size of the conductive phase diminishes during the removing process and the percolation transition occurs between a percolating and a non-percolating state.

In previous work \cite{JSTAT2,JSTAT8,PRE16,JSTAT9}, we studied the problem of inverse percolation by removing linear objects from two-dimensional (2D) lattices. This corresponds to the complementary form of the standard percolation of straight rigid rods on a discrete lattice \cite{Vandewalle,Kondrat,Lebovka,Tara2012,EPJB1,Budi2012,Slutskii} and is conceptually similar to the void continuum percolation problem or Swiss-cheese percolation \cite{Kertesz,Kerstein,vanderMarck,Klemm,Rintoul2000,Yi2006,Yi2012,Priour2018}. In this case, one considers a system of overlapping objects and asks when the space not occupied by the objects percolates. The problem is very similar to the original definition involving fluid flow through a porous media \cite{Sahimi}.

In Ref. \cite{JSTAT2}, the problem of removing linear site $k$-mers (particles occupying $k$ consecutive sites along one of the lattice directions) from square lattices was studied by numerical simulations and finite-size analysis. The percolating phase occurring at high concentrations is separated from a non-percolating phase by a continuous transition occurring at an intermediate critical density $p_{c,k}$. This critical density was calculated as a function of $k$. The results, obtained for $k$ ranging from 2 to 256, showed a nonmonotonic size $k$ dependence for $p_{c,k}$, which rapidly decreases for small particles sizes ($1 \leq k \leq 3$). Then, it grows for $k=4$, 5 and 6, goes through a maximum at $k = 7$, and finally decreases again and asymptotically converges towards a definite value for large values of $k$ [$p_{c,k \rightarrow \infty}=0.454(4)$]. A similar study for triangular lattices was carried out in Ref. \cite{JSTAT8}. In this case, the maximum occurs at $k = 11$ and the convergence value is $p_{c, k \rightarrow \infty}=0.506(2)$.

In terms of network robustness, the results discussed in the paragraph above indicate that, for large $k$-mers ($k \geq 7$ for square lattices and $k \geq 11$ for triangular lattices) and a same fraction of removed sites (or attack), the robustness of the network increases with the attack size ($k$). These findings are consistent with those from Refs. \cite{Kornbluth,Lowinger}, where the vulnerability of networks during the process of cascading failures was investigated. The authors studied the influence of the characteristics of the initial attack on the vulnerability of the networks, showing that random attacks on single nodes are much more effective than correlated attacks on groups of close nodes.

More recently, numerical simulations and finite-size scaling analysis have been carried out to study the problem of inverse bond percolation by removing linear bond $k$-mers (objects formed by $k$ consecutive bonds along one of the lattice directions) from square lattices \cite{PRE16}. The obtained results showed that the inverse percolation threshold is a decreasing function of $k$ in the range $1 \leq k \leq 18$. For $k > 18$, all jammed configurations are percolating states, and consequently, there is no nonpercolating phase. As in previous cases \cite{JSTAT2,JSTAT8}, the decreasing behavior of the inverse percolation threshold as a function of $k$ clearly indicates that random attacks on single nodes ($k = 1$) are much more effective than correlated attacks on groups of close nodes. In addition, the loss of the phase transition has very interesting implications in terms of network attacks. In fact, for large $k$-mers ($k > 18$), the lattice remains connected even when the highest allowed concentration of removed bonds is reached.

A similar behaviour was observed for inverse site percolation of linear $k$-mers in the presence of impurities \cite{JSTAT9}. As in Ref. \cite{PRE16}, the percolation phase transition disappears for a certain value of $k$, which depends on the value of the fraction of impurities. The study complements previous work in homogeneous lattices \cite{JSTAT2,JSTAT8,PRE16}, revealing that the construction of networks with low local connectivity (or low clustering capacity), as occurs in the model for increasing values of the fraction of impurities, is an effective strategy against correlated attacks on groups of close nodes (large $k$'s).

In the case of void percolation, the effect of the shape of the removed objects has been widely studied \cite{Rintoul2000,Yi2006,Yi2012,Priour2018}. The same has not happened for inverse percolation on discrete lattices, where most of the attention has been devoted to the removal of linear clusters of sites (bonds) \cite{JSTAT2,JSTAT8,PRE16,JSTAT9}.

The aim of the present work is to extend previous studies to the removal of more compact objects such as $k \times k$ square tiles (or $k^2$-mers). For this purpose, extensive numerical simulations supplemented by analysis using finite-size scaling theory have been carried out to study the problem of inverse percolation by removing $k^2$-mers from square lattices. Our interest is in investigating the effect of the shape of the removed object (structure of the attack) on the connectivity properties of the damaged lattice.

It is also interesting to compare the results obtained for inverse percolation with those reported for the standard percolation problem of $k^2$-mers on square lattices, where the percolation phase transition disappears for $k \geq 4$ \cite{Nakamura86,Nakamura,PRE19}.

The paper is organized as it follows: the model is presented in Section \ref{modelo}. Jamming and percolation properties are studied in Section \ref{perco}. Finally, the conclusions are drawn in Section \ref{conclu}.

\section{The model}\label{modelo}

Let us consider a square lattice of $M= L \times L$ sites that represents our surface. Each site of the lattice only has two possible states of occupation: empty or occupied. Nearest-neighbor occupied sites form structures called clusters. The distribution of these occupied sites determines the probability of the existence of a large cluster (also called ``infinite" cluster, inspired by the thermodynamic limit) that connects from one side of the lattice to the other.

As it was already mentioned, the central idea of the inverse percolation model is based on removing objects from an initial configuration where all sites are occupied and finding the maximum concentration $p$ for which the connectivity disappears. We called this particular value of the concentration as \textit{inverse percolation threshold} \cite{JSTAT2,JSTAT8,PRE16,JSTAT9}. In this study, the removed species are square tiles composed by $k \times k$ occupied sites. Accordingly, the inverse percolation threshold will be denoted as $p_{c,k}$.

The dilution of the lattice with $k^{2}$-mers is carried out following a conventional RSA process \cite{Feder,Evans,Talbot} and considering periodic boundary conditions in both lattice directions. The scheme consists of three steps, namely, (i) starting from an initially fully occupied lattice; (ii) then, a square tile of $k \times k$ sites is chosen randomly and if those sites are occupied, a $k^{2}$-mer is removed; otherwise, the attempt is rejected; (iii) steps $(i)-(ii)$ are repeated until a desired concentration $p=1 - k^{2}N/M$ is reached ($N$ is the number of the removed $k^{2}$-mers).

Figure 1 shows a typical lattice configuration after removal of $2 \times 2$ tiles (open circles joined by lines) from the two-dimensional square lattice. The solid circles represent the occupied circles.

\begin{figure}
\begin{center}
\includegraphics[width=0.6\columnwidth]{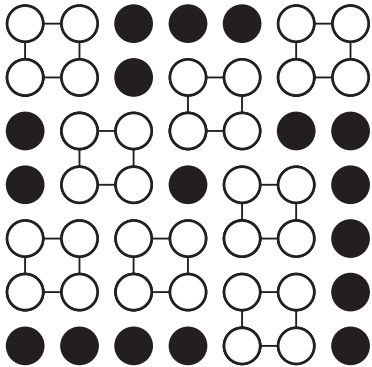}
\caption{Schematic representation of a square lattice in which some  $2 \times 2$ tiles (open circles joined by solid lines) have been removed. The solid circles represent the occupied sites.\label{Fig-esquema}}
\end{center}

\end{figure}

\section{INVERSE PERCOLATION AND JAMMING PROPERTIES}\label{perco}

The inverse percolation problem results quite simple for the case of removing single sites or bonds, when the inverse and standard problems are symmetrical. However, if some sort of correlation exists, as in the case of removing tiles of $k \times k$  elements, the statistical problem becomes exceedingly difficult and the percolation threshold has to be estimated numerically by means of computer simulations.

\subsection{Jamming coverage}

Let us consider the complementary lattice to the original lattice, where each empty (occupied) site of the original lattice transforms into a occupied (empty) one of the complementary lattice. Under these conditions, the filling process in the complementary lattice (dilution process in the original lattice) is equivalent to a RSA process of $k^2$-mers. Accordingly, both problems share formal aspects, terminology and essential characteristics, as the existence of a nontrivial state, the jammed saturation state.

In Fig. \ref{Fig-esquema} it can be easily seen that the geometry of the $k^{2}$-mer excludes the possibility of continuing to eliminate tiles even if there are occupied sites on the lattice. Thus, the jamming coverage $p_{j,k}$ (the subindex $k$ indicates that the jamming coverage was reached after removing square tiles of side $k$) is the concentration of occupied sites at which no more objects can be removed from the lattice (the lattice is blocked). From the relationship between original and complementary lattices, it is straightforward that $p_{j,k} = 1 - p'_{j,k}$, where $p'_{j,k}$ is the jamming coverage corresponding to a standard RSA process of $k^2$-mers on square lattices. The dependence of $p'_{j,k}$ as a function of the size $k$ has recently been studied \cite{PRE19}. In Ref. \cite{PRE19}, numerical simulations were performed for $k$ in the range 2-100, and several values of $L/k: 128, 192, 256, 320, 384$, and $448$. For each $k-L$ pair, the results were obtained by averaging on $2 \times 10^5$ independent samples. The authors found that (1) $p'_{j,k}$ is a decreasing function of $k$, and (2) the best fit to $p'_{j,k}$ (obtained for $k \geq 2$) corresponds to the expression: $p'_{j,k}$=$A+B/k+C/k^{2}$, being $A=p'_{j,k=\infty}$=0.5623(3), $B$=0.3098(2) and $C$=0.1277(2). Then, the fraction of occupied sites ranges from 1 to $p_{j,k}$, where
\begin{eqnarray}\label{pjk}
p_{j,k} & = & 1 - p'_{j,k} \nonumber \\
& = & 0.4377 -\frac{0.3098}{k}-\frac{0.1277}{k^2}.
\end{eqnarray}

\begin{figure}
 \includegraphics[scale=0.5]{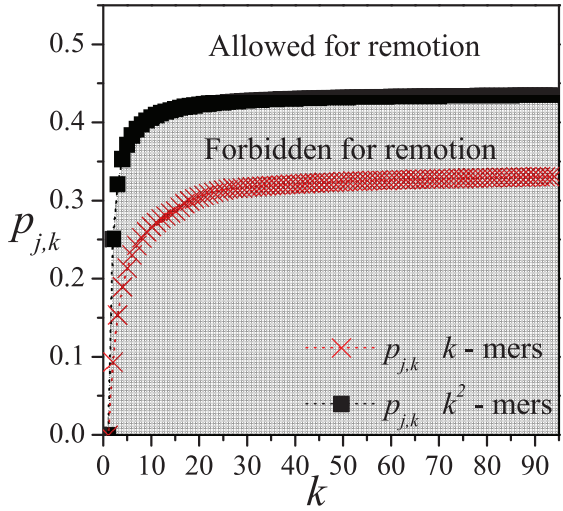}
\caption{ Curves of $p_{j,k}$ vs $k$ for tiles (line and solid squares) and linear $k$-mers (line and crosses). The grey zone represents the space of the concentrations that are not possible to access for the remotion of tiles because of the blocking of the lattice.\label{figJam}}
\end{figure}

The jamming curve in Eq. (\ref{pjk}) is shown in Fig. \ref{figJam} (line and solid squares). For comparison, the figure also includes the jamming curve corresponding to the problem of removing straight rigid $k$-mers from square lattices (line and crosses): $p_{j,k}= 0.34 - 1.071/k + 3.47/k^2 (k \geq 48)$ \cite{JSTAT2,Bonnier}. This expression was obtained by fitting simulation data for segments of length $k$ between 2 and 512 and lattices of linear size $L$ between 128 and 4096 \cite{Bonnier}.

The space of the parameter $p$ is separated in two regions by the jamming curve. The region above the curve of $p_{j,k}$ represents the space of all the allowed values of $p$ (values of $p$ which can be reached by removing objects from the surface). On the other hand, the region below to the curve of $p_{j,k}$ corresponds to a forbidden region of the space. The above means that if we started from a fully occupied lattice ($p$=1), we can remove components while the concentration is higher than $p_{j,k}$ but we cannot access to values of $p$ so that $p < p_{j,k}$. In the figure, the grey zone indicates the space of the concentrations that are not possible to access by removing $k \times k$ tiles from square lattices.

Clearly, the probability of blocking the lattice depends both on the geometry of the surface and on the shape of the removed objects. As it can be visualized from Fig. \ref{figJam}, the $k$-mers jamming curve remains below the corresponding $k^2$-mers curve, indicating that the lattice is blocked at higher concentrations of occupancy for tiles than for rigid rods. In terms of RSA process (in the complementary lattice), it means that linear $k$-mers are more effective in filling the lattice than $k \times k$ square tiles.

It is important to note that the term ``jamming", in the sense used in the present paper, refers to the final state generated by irreversible adsorption of structured objects (in this case RSA), in which no more objects can be deposited due to the absence of free space of appropriate size and shape \cite{Feder,Evans,Talbot}. This phenomenon should not be confused with the classical jamming transition from a flowing to a rigid state, which is a paradigm for thinking about how many different types of fluids (from molecular liquids to macroscopic granular matter) develop rigidity \cite{Liu2010}.

However, some critical properties have been identified in relation to the jamming phenomenon associated with the RSA problem. To understand this point, it is convenient to define the jamming probability $W_L(p)$ \cite{PHYSA38}. In our case, $W_L(p)$ can be defined as the probability that a $L \times L$ lattice reaches a coverage $p$ by removing sets of particles of size $k \times k$ ($k^2$-mers). The procedure to determine $W_L(p)$ consists of the following steps: (a) the construction of the lattice (initially fully occupied) and (b) the removal of objects on the lattice up to the jamming limit $p_{j,k}$. In the late step, the quantity $m_i(p)$ is calculated as
\begin{equation}\label{mi}
m_i(p)=\left\{
\begin{array}{cc}
1 & {\rm for}\ \ p \geq p_{j,k} \\
0  & {\rm for}\ \ p < p_{j,k} .
\end{array}
\right.
\end{equation}
$n$ runs of such two steps (a)-(b) are carried out for obtaining the number $m(p)$ of them for which a lattice reaches a coverage $p$,
\begin{equation}\label{m}
m(p) = \sum_{i=1}^n m_i(p).
\end{equation}
Then, $W_L(p)= m(p)/n$ is defined and the procedure is repeated for different values of $L$.

\begin{figure}
\begin{center}
\includegraphics[scale=0.5]{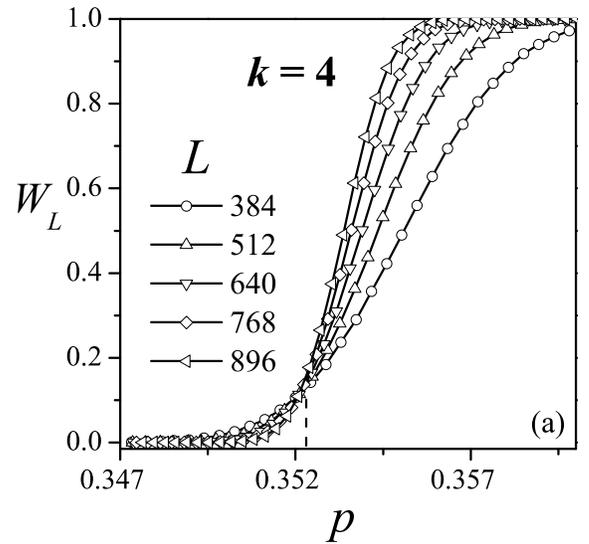}
\includegraphics[scale=0.5]{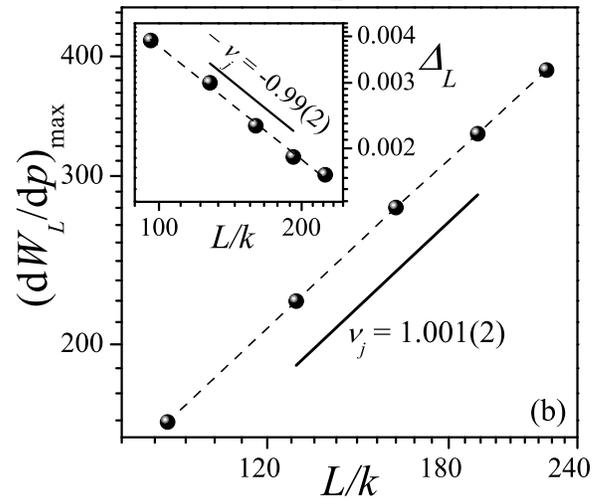}
\includegraphics[scale=0.5]{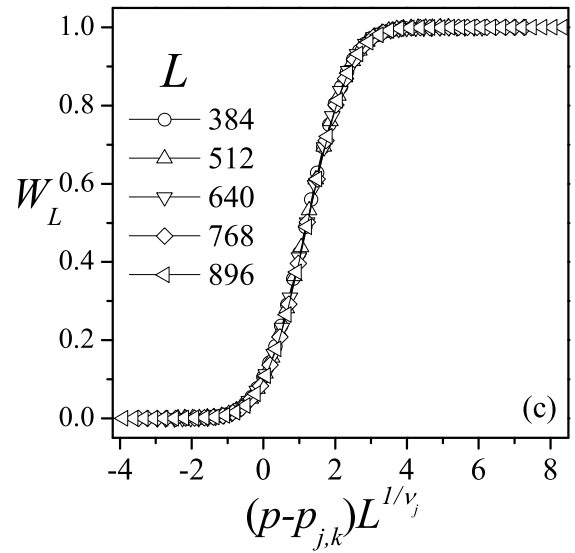}
\caption{(a) Curves of the jamming probability $W_L(p)$ as a function of the fraction of occupied sites $p$ for $k=4$ and different lattice sizes as indicated. (b) Log-log plots of $(\mathrm{d}W_{L}/\mathrm{d}p)_{\rm max}$ and $\Delta_{L}$ as a function of $L$ for the case shown in part (a). According to Eq. (\ref{derivada}) the slope of each line corresponds to 1/$\nu_{j}$ (or to -1/$\nu_{j}$ in the case of Eq. (\ref{Delta})). (c) Data collapse of the jamming probability, $W_L$ versus $\left( p - p_{j,k} \right) L^{1/\nu_j}$ for the data in part (a). The curves were obtained using $p_{j,k=4}=0.35207$ \cite{PRE19} and $\nu_j=1$. 
\label{FigWL}}
\end{center}
\end{figure}

Figure \ref{FigWL}(a) shows the behavior of $W_L(p)$ for $k=4$ and different values of the lattice size $L=384, 512, 640, 768, 896$. During the removing process, the fraction of particles on the lattice diminishes (the $p$-axis varies between 1 and $p_{j,k}$). As it can be observed from the figure, $W_L(p)$ varies continuously between 1 and 0, with a sharp fall around $p_{j,k}$. Even when the probabilities show a dependence on the system size, $W_L(p)$ is independent of the system size for $p=p_{j,k}$ \cite{PHYSA38}. Thus, the value of $p_{j,k}$  can be obtained from the crossing point of the curves of $W_L(p)$ for different lattice sizes.

The jamming probability can also be used to determine the jamming exponent $\nu_j$. As established in the literature \cite{Vandewalle}, $\mathrm{d} W_L(p)/\mathrm{d} p$ can be fitted by the Gaussian function. This is a good approximation for the purpose of locating its maximum. Then, according to the finite-size scaling theory \cite{Vandewalle}, the maximum of the derivative of the jamming probability $[\mathrm{d} W_L(p)/\mathrm{d} p]_{max}$ and the width of the transition $\Delta_L$ behave asymptotically as
\begin{equation}\label{derivada}
  \left(\frac{\mathrm{d}W_L}{\mathrm{d} p}\right)_{\rm max}\propto L^{1/\nu_j},
\end{equation}
and
\begin{equation}\label{Delta}
  \Delta_{L}\propto L^{-1/\nu_j}.
\end{equation}

Figure \ref{FigWL}(b) shows, in a log-log scale, $({\mathrm{d}W_{L}}/{\mathrm{d}p})_{\rm max}$ and $\Delta_{L}$ (inset) as a function of $L$ for $k$=4, where $\nu_j$ can be obtained from the inverse of the slopes of the lines that fit the data. In this case $\nu_j$=1.001(2) (main figure) and $\nu_j$=0.99(2) (inset). The study was repeated for other sizes $k$. In all cases, the obtained values of $\nu_j$ remain close to 1. This finding confirms recent investigations on RSA processes on Euclidean lattices \cite{RSACWI}. The results in Ref. \cite{RSACWI} showed that $\nu_j=2/d$, where $d$ is the dimensionality of the lattice. The values of $\nu_j$ do not depend on size and shape of the depositing objects.

As shown in Figs. \ref{FigWL}(a) and \ref{FigWL}(b), the properties of $W_L(p)$ are identical to those of $R^X_L(p)$ in standard percolation transitions (this probability will be discussed in details in the next section). Namely, $R^X_L(p)$ obeys the same scaling relation in Eqs. (\ref{derivada}) and (\ref{Delta}), and the intersection of the curves of $R^X_L(p)$ for different system sizes can be used to determine the critical point that characterizes the phase transition occurring in the system. Then, based on these features, we propose the following scaling behavior at criticality for the jamming probability:
\begin{equation}\label{WLcol}
W(p)=  \overline{W}\left[\left( p - p_{j,k} \right) L^{1/\nu_j}\right],
\end{equation}
where $\overline{W}$ is the corresponding scaling function.

The scaling tendency in Eq. (\ref{WLcol}) has been tested by plotting $W_L(p)$ versus $(p-p_{j,k})L^{1/\nu_j}$ and looking for data collapsing. As an example, Fig. \ref{FigWL}(c) shows the obtained results for $k=4$. Using the values of $p_{j,k=4}=0.35207$ \cite{PRE19} and $\nu_j=1$, the curves present an excellent scaling collapse. This data collapse study, which has been reported for the first time in the literature for the jamming probability $W_L(p)$, allows for consistency check of the value $\nu_j=1$ calculated in Fig. \ref{FigWL}(b).

\subsection{Percolation threshold}

Once the limiting parameters $p_{j,k}$'s are determined, we will focus on finding the phase diagram given by the evolution of the inverse percolation threshold with the size of the removed tiles. The percolation transition is analogous to continuous phase transitions that occur in thermodynamic systems and, as is known, a phase transition can only take place in the thermodynamic limit (this is $N  \rightarrow \infty$ $M  \rightarrow \infty$ and $N/M$ (finite)). In finite systems (as the ones that are possible to simulate computationally no matter how large $N$ and $M$ are) is not possible to have a sharply defined threshold and is the finite-size scaling theory the one that sets up the basis to achieve the percolation threshold of the system with accuracy. The results presented here were derived through simulations complemented with finite-size scaling analysis.

The main information is obtained from the probability $R^X_{L,k}(p)$ that a lattice composed of $L \times L$ sites percolates at the concentration $p$ after the removal of $k \times k$ tiles \cite{Stauffer}. The index $X$ in the definition of $R$ may have the following meanings:
\begin{itemize}
    \item $R^{U}_{L,k}(p)$: the probability of finding a cluster which percolates on any direction ($x$-direction or $y$-direction),
  \item $R^{I}_{L,k}(p)$: the probability of finding a cluster which percolates in the two (mutually perpendicular) directions ($x$-direction and $y$-direction),
  \item $R^{A}_{L,k}(p)$=$\frac{1}{2}[R^{U}_{L,k}(p)+R^{I}_{L,k}(p)]$.
\end{itemize}

Basically, each simulation run consists of the following steps: $(i)$ the construction of a lattice of linear size $L$ with a coverage $p$ according to the dilution procedure described in Sec. \ref{modelo}, and $(ii)$ the cluster analysis using the Hoshen and Kopelman algorithm \cite{Hoshen}. In this last step, the size of the largest cluster $S_L$ is determined, as well as the existence of a percolating island. We consider open boundary conditions for the percolation calculations.

$m_L$ independent runs of such two steps procedure were carried out for each lattice size $L$ and concentration $p$. Then, $R^X_{L,k}(p)$ was defined as the ratio between the runs that presented a percolation cluster, $m^X_L$, and the total attempts $m_L$. So, $R^X_{L,k}(p) = m^X_L/m_L$ is defined for the desired criterion among $X=\{I,U,A\}$ and the procedure is repeated for different values of $L$, $p$ and $k \times k$ size of the tiles. For each value of $k$, $m_L = 10^5$ independent random samples were carried out with $L/k = 128, 256, 320, 384, 448$, and $512$. As it can be appreciated, this represents extensive calculations from the computational point of view. Then, the finite-scaling theory can be used to determine the percolation threshold and the critical exponents with reasonable accuracy \cite{Stauffer,Yone1,Binder}.

\begin{figure}
\includegraphics[scale=0.5]{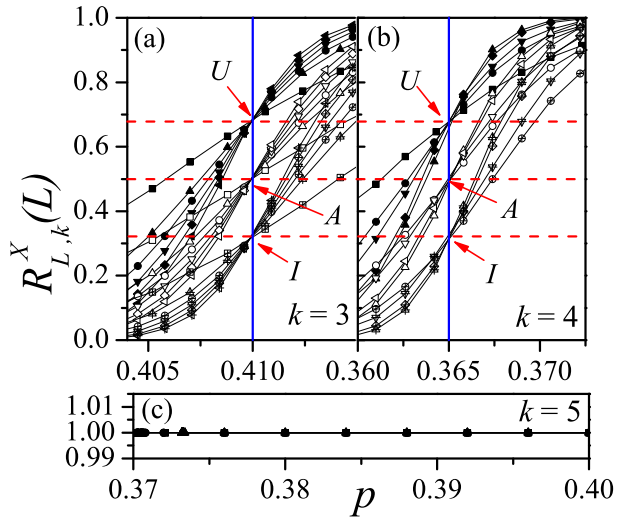}
\caption{Fraction of percolating lattices $R^{X}_{L,k}(p)$ ($X=\{I,U,A\}$, as indicated) as a function of the concentration $p$ for $k = 3$ (a), $k = 4$ (b), $k = 5$ (c), and different lattice sizes: $L/k = 128$, squares; $L/k = 256$, circles; $L/k = 320$, up triangles; $L/k = 384$, down triangles; $L/k = 448$, left triangles; and $L/k = 512$, right triangles. The statistical errors are smaller than the symbol sizes. \label{GraphR}}
\end{figure}

The probability curves $R^X_{L,k}(p)$ are shown in Fig. \ref{GraphR} for $k=3$ (a), $k=4$ (b), $k=5$ (c), and different values of $L/k$. From parts (a-b), it is observed that the $y$-axis values of the crossing points ($R^{X^*}$) depend on the criterion $X$ used: $R^{A^*} \approx 0.50$, $R^{I^*} \approx 0.32$ and $R^{U^*} \approx 0.68$.  These results coincide (within the numerical errors) with the corresponding exact values for standard percolation: $A$ criterion, $1/2$ \cite{Cardy,Simmons}; $I$ criterion, $0.32212045\dots$ \cite{Simmons,Watts} and $U$ criterion, $0.67788954\dots$ \cite{Simmons,Watts}. In addition, the crossing points do not modify their numerical value for the different sizes studied ($k=1,2,3,4$). This finding represents a first indication that the universality class of the phase transition involved in the problem is conserved no matter the values of $k$.

The situation is different for $k = 5$ [Fig. \ref{GraphR}(c)], where the curves of $R^{X}_{L,k}(p)$ remain around 1, demonstrating there is only one phase (the percolating phase) in the whole range of allowed values of $p$ (there is not phase transition). This finding indicates that the percolation phase transition disappears for $k > 4$. In other words, as $k^2$-mers with $k > 4$ are removed from a square lattice, the jamming transition occurs before the percolating island can be separated into a finite number of isolated clusters.

As mentioned, the percolation phase transition is well defined by its critical exponents.  At this point, we are capable of finding the critical exponent of the correlation length, $\nu$, from numerical data. This exponent is of importance because it is necessary in order to calculate the percolation threshold. The finite-size scaling theory allows to estimate  $\nu$ through different methods. One route is from the maximum of the derivative of the functions $R^X_{L,k}(p)$ \cite{Stauffer},
\begin{equation}
\left(\frac{d R^{X}_{L, k}}{d p} \right)_{\rm max} \propto L^{1/\nu}. \label{lambda}
\end{equation}

In order to apply Eq. (\ref{lambda}), it is convenient to fit $R^{X}_{L,k}(p)$ with some approximating function through the least-squares method. This allows us to express $R^{X}_{L,k}(p)$
as a function of continuous values of $p$. The fitting curve is the {\em error function} because $dR^{X}_{L,k}(p)/dp$ is expected to behave approximately like the Gaussian distribution \cite{Yone1}. We use the term approximately because the behavior of $dR^{X}_{L,k}(p)/dp$ is known not to be a Gaussian in all range of coverage \cite{Newman3}. However, this quantity is approximately Gaussian near the peak, and fitting with a Gaussian function is a good approximation for the purpose of locating its maximum. Thus,
\begin{equation}\label{ecu1}
    \frac{dR^{X}_{L,k}}{dp}=\frac{1}{\sqrt{2\pi}\Delta^{X}_{L,k}}\exp \left\{ -\frac{1}{2} \left[\frac{p-p_{c, k}^{X}(L)}{\Delta^{X}_{L, k}}
    \right]^2 \right\},
\end{equation}
where $p_{c, k}^{X}(L)$ is the concentration at which the slope of $R^{X}_{L,k}(p)$ is the largest and $\Delta^{X}_{L,k}$ is the standard deviation from $p_{c, k}^{X}(L)$.

\begin{figure}
\includegraphics[scale=0.5]{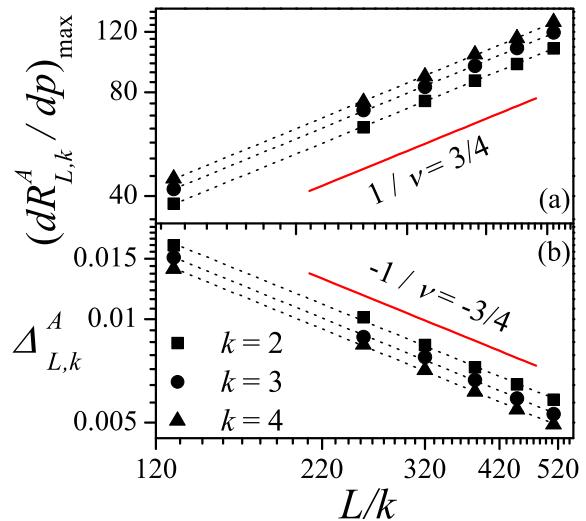}
\caption{(a) Log-log plot of $\left(d R^{A}_{L, k}/d p \right)_{\rm max}$ as a function of $L/k$ for $k=2$ (squares), $k=3$ (circles) and $k=4$ (triangles). According to Eq. (\ref{lambda}) the slope of each line corresponds to $1/ \nu=3/4$. (b) $\ln \left( \Delta_{L,k}^A \right)$ as a function of $L/k$ for $k=2$ (squares), $k=3$ (circles) and $k=4$ (triangles). According to Eq. (\ref{delta}), the slope of each curve corresponds to $-1/\nu=-3/4$. \label{Graphnu}}
\end{figure}

In Fig. \ref{Graphnu}(a), $\ln\left[\left(d R^{A}_{L, k}/d p \right)_{\rm max}\right]$ has been plotted as a function of $\ln\left[ L/k \right]$ (note the log-log functional dependence) for $k$ = 2, 3 and 4. According to Eq. (\ref{lambda}) the slope corresponds to $1/ \nu$.

Another alternative way to obtain  $\nu$ is given by the divergence of the root mean square deviation of the threshold observed from their average values, $\Delta_{L,k}^A$ in Eq. (\ref{ecu1})
\cite{Stauffer},
\begin{equation}
\Delta_{L,k}^X \propto L^{-1/\nu}. \label{delta}
\end{equation}
Figure \ref{Graphnu}(b) shows $\ln \left(\Delta_{L,k}^A \right)$ as a function of $\ln(L/k)$ (note the log-log functional dependence)  for $k$ = 2, 3 and 4. According to Eq. (\ref{delta}), the slope corresponds to $-1/ \nu$.

For both methods, the values of $1/ \nu$ remain constant and close to $3/4$. The study in Fig.  \ref{Graphnu} was repeated for the $I$ and $U$ percolation criteria. In all cases, the results coincide, within numerical errors, with the exact value of the critical exponent of the ordinary percolation $\nu=4/3$.

Once $\nu$ was determined and with previous values of $p_{c, k}^{X}(L)$ [Eq. (\ref{ecu1})], a scaling analysis can be done to determine the percolation threshold in the thermodynamic limit \cite{Stauffer}. Thus, we have
\begin{equation}\label{extrapolation}
p_{c,k}^{X}(L)= p_{c,k}^{X}(\infty) + A^X L^{-1/\nu},
\end{equation}
where $A^X$ is a non-universal constant.

\begin{figure}
\includegraphics[scale=0.5]{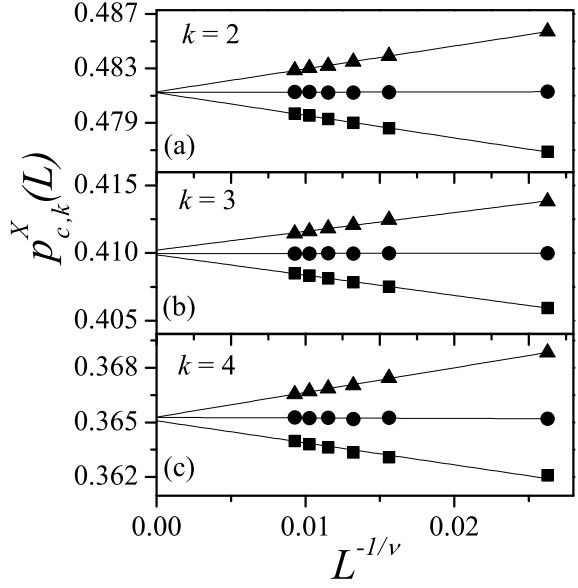}
\caption{Extrapolation of the percolation threshold for an $L$-lattice $p_{c, k}^{X}(L)$ $(X=\{I,U,A\})$ towards the thermodynamic limit according to the theoretical prediction given by Eq. (\ref{extrapolation}) for the data in Fig. \ref{GraphR}: (a) $k=2$; (b) $k=3$ and (c) $k=4$. Triangles, circles and squares denote the values of $p_{c, k}^{X}(L)$ obtained by using the criteria $I$, $A$ and $U$, respectively. The bar error in each measurement is smaller than the size of the corresponding symbol.\label{fig5}}
\end{figure}

Figure \ref{fig5} shows the extrapolation towards the thermodynamic limit of $p_{c, k}^{X}(L)$ ($X = {I,U,A}$ and $k=3,4$) according to Eq.(\ref{extrapolation}). Combining the three estimates for each size $k$, the final values of $p_{c,k}(\infty)$ can be obtained. Additionally, the maximum of the differences between $\mid p^{U}_{c,k} - p^{A}_{c,k}\mid$ and $\mid p^{I}_{c,k} - p^{A}_{c,k}\mid$
gives the error bar for each determination of $p_{c,k}(\infty)$. The values obtained in Fig. \ref{fig5} were: $p_{c,k=2}(\infty)=0.48115(5)$, $p_{c,k=3}(\infty)=0.40997(9)$, and $p_{c,k=4}(\infty)=0.36500(11)$. For the rest of the paper, we will denote the percolation threshold for each size $k$ by $p_{c,k}$ [for simplicity we will drop the symbol``$(\infty)$"].

\begin{figure}
\includegraphics[scale=0.5]{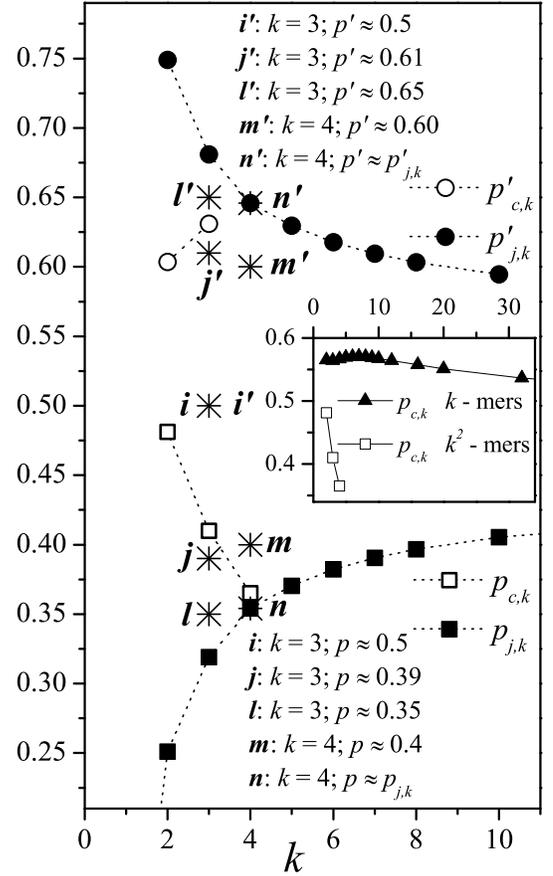}
\caption{Inverse percolation threshold $p_{c, k}$ (open squares) and jamming coverage $p_{j, k}$ (solid squares) as a function of $k$ for the problem of removing $k^2$-mers from square lattices. Percolation threshold $p'_{c, k}$ (open circles) and jamming coverage $p'_{j, k}$ (solid circles) as a function of $k$ for the standard problem of depositing $k^2$-mers on square lattices (data correspond to results reported in Ref. \cite{PRE19}). Points $i,j,l,m,n$ and $i',j',l',m',n'$ are explained in the text. Inset: Comparison between the inverse percolation thresholds obtained by removing $k^2$-mers from square lattices (open squares) and the corresponding ones obtained by removing linear $k$-mers from square lattices (solid triangles) \cite{JSTAT2}.\label{fig6}}
\end{figure}

\begin{figure}
\begin{center}
\includegraphics[scale=0.5]{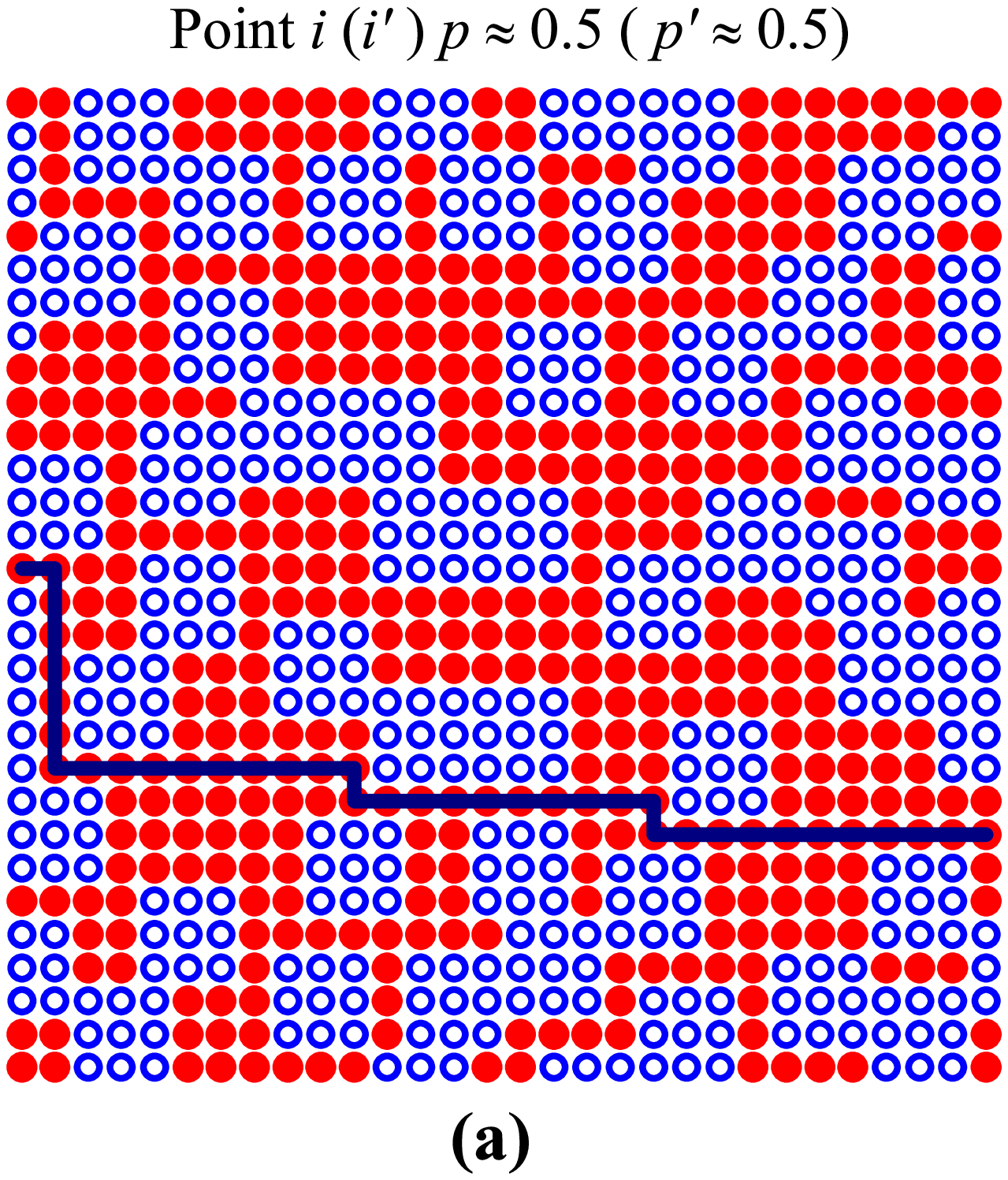}
\includegraphics[scale=0.5]{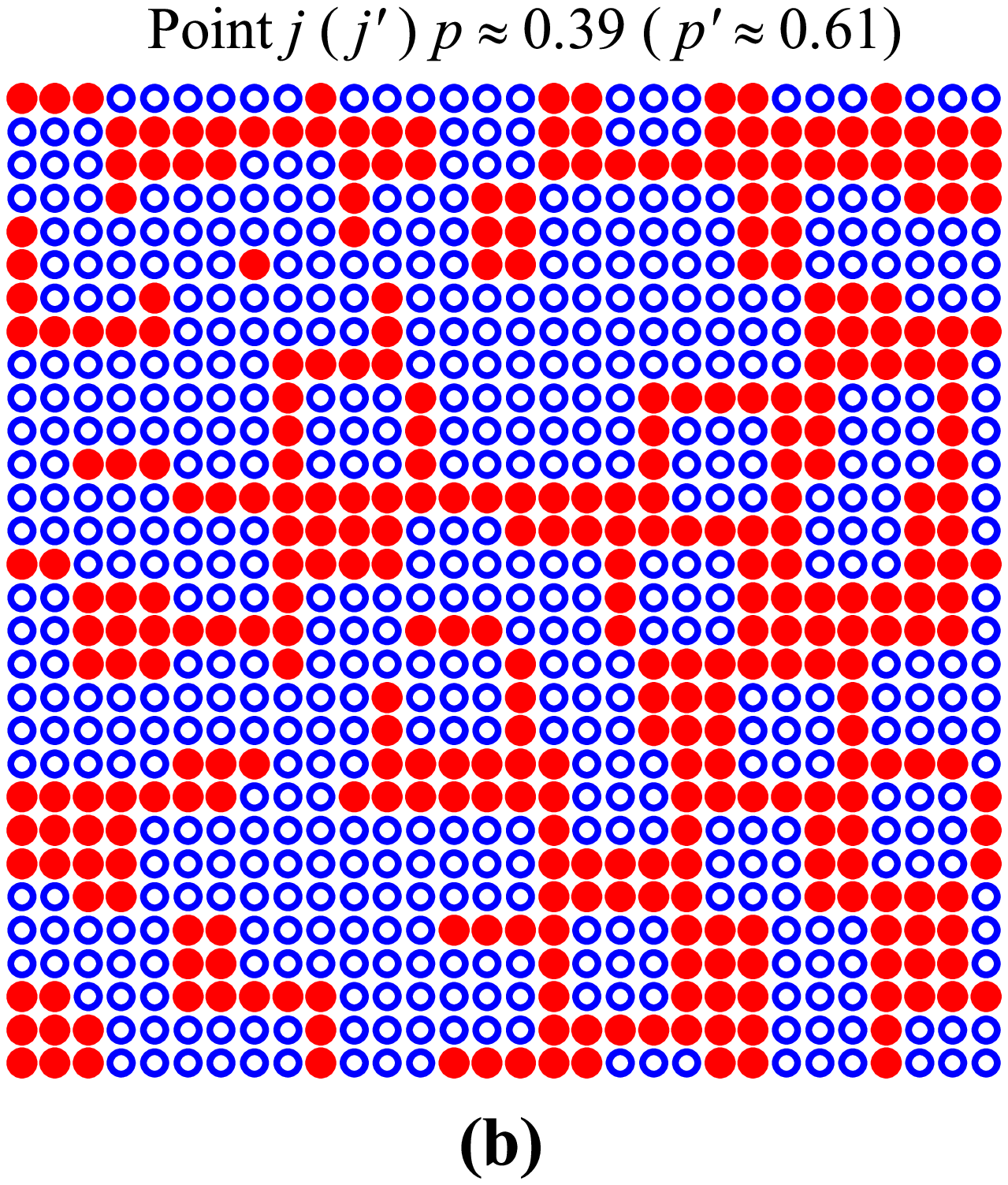}
\includegraphics[scale=0.5]{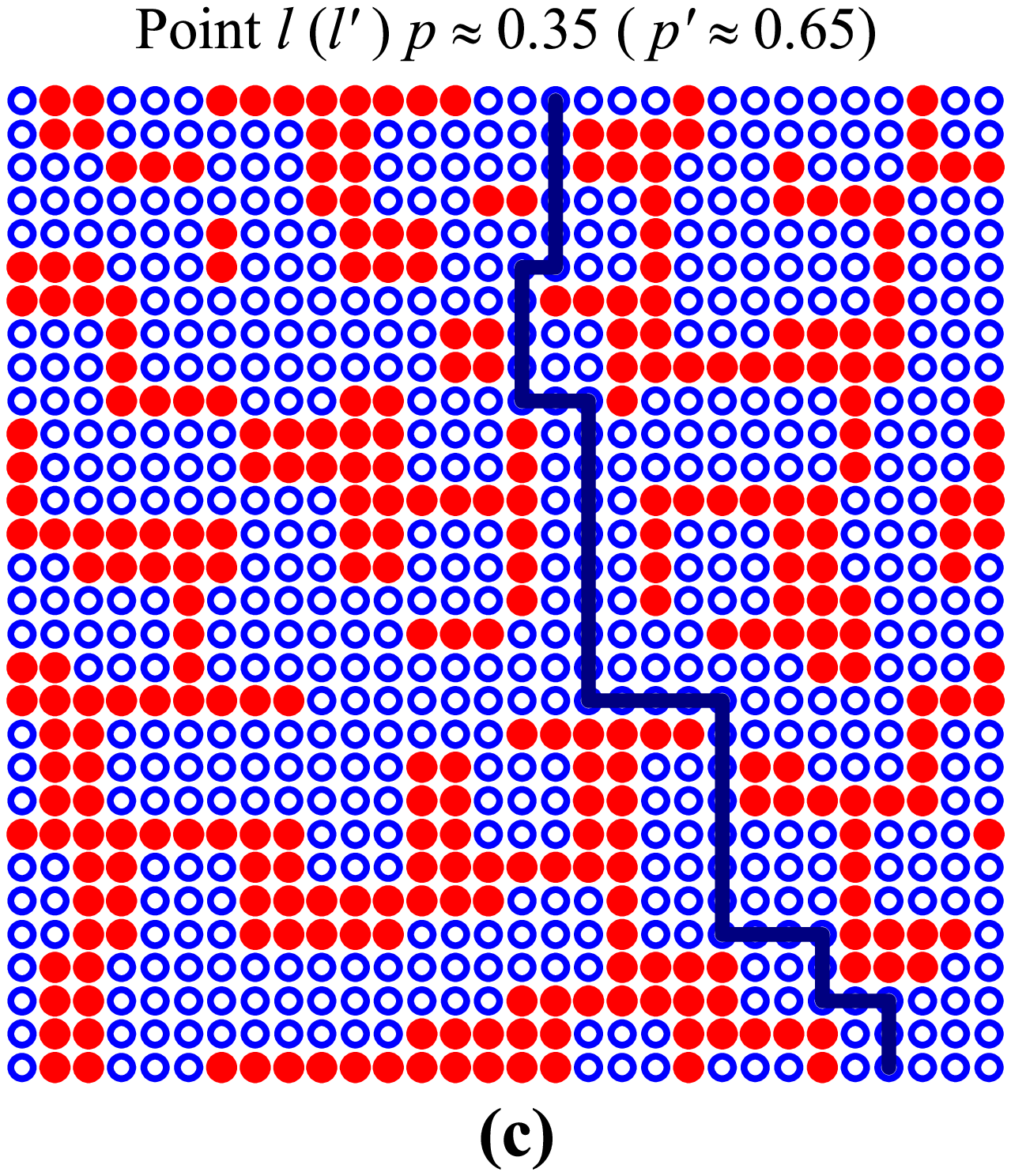}
\caption{(a) Snapshot of a typical lattice configuration in points $i$ and $i'$ of Fig. \ref{fig6}. (b) Same as part (a) for the points $j$ and $j'$ of Fig. \ref{fig6}. (c) Same as part (a) for the points $l$ and $l'$ of Fig. \ref{fig6}. The meaning of red solid circles and blue open circles is given in the text. Thick line denotes a percolation path connecting two opposite sides of the lattice.
\label{Figsnapk3}}
\end{center}
\end{figure}

\begin{figure}
\begin{center}
\includegraphics[scale=0.5]{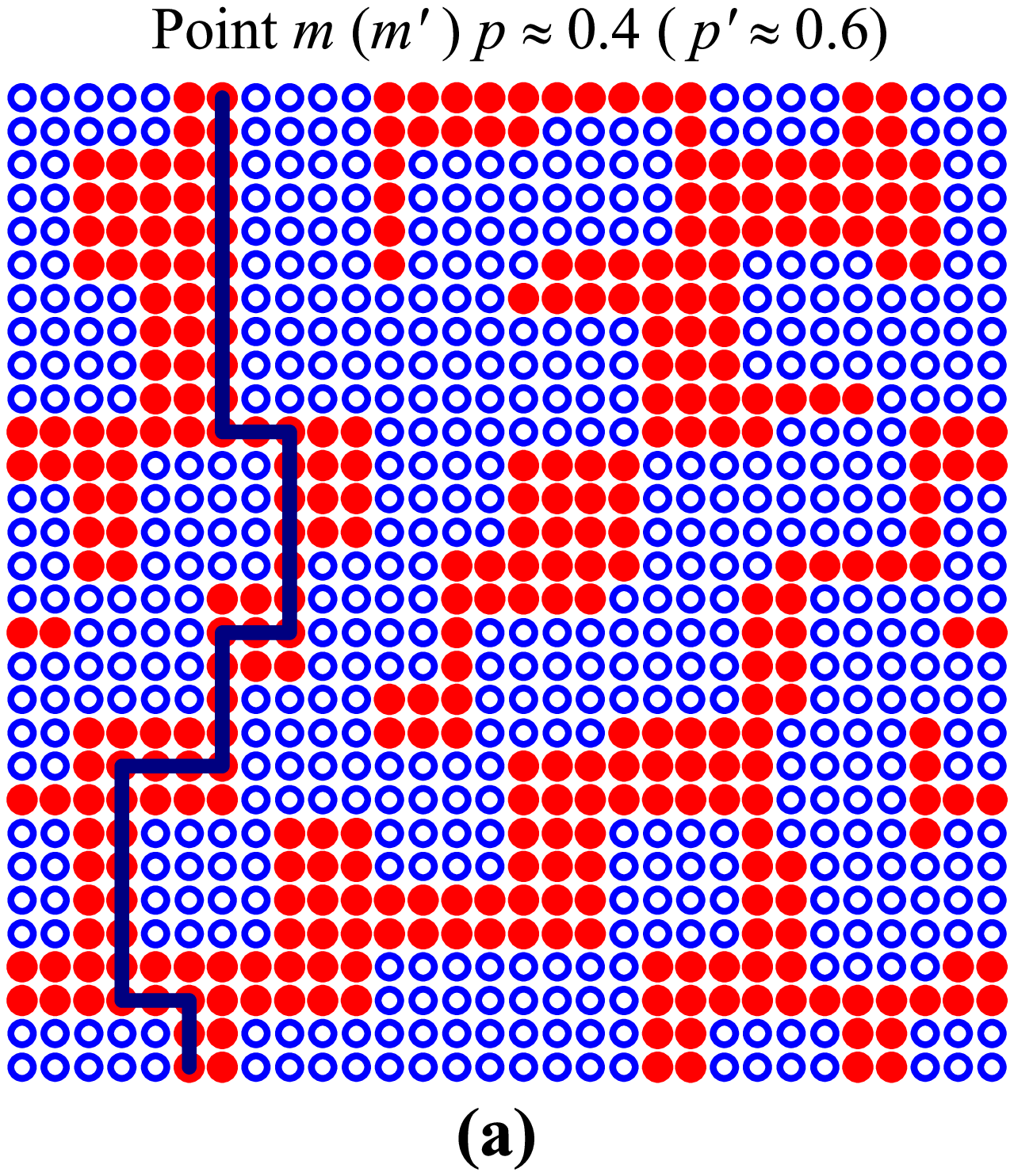}
\includegraphics[scale=0.5]{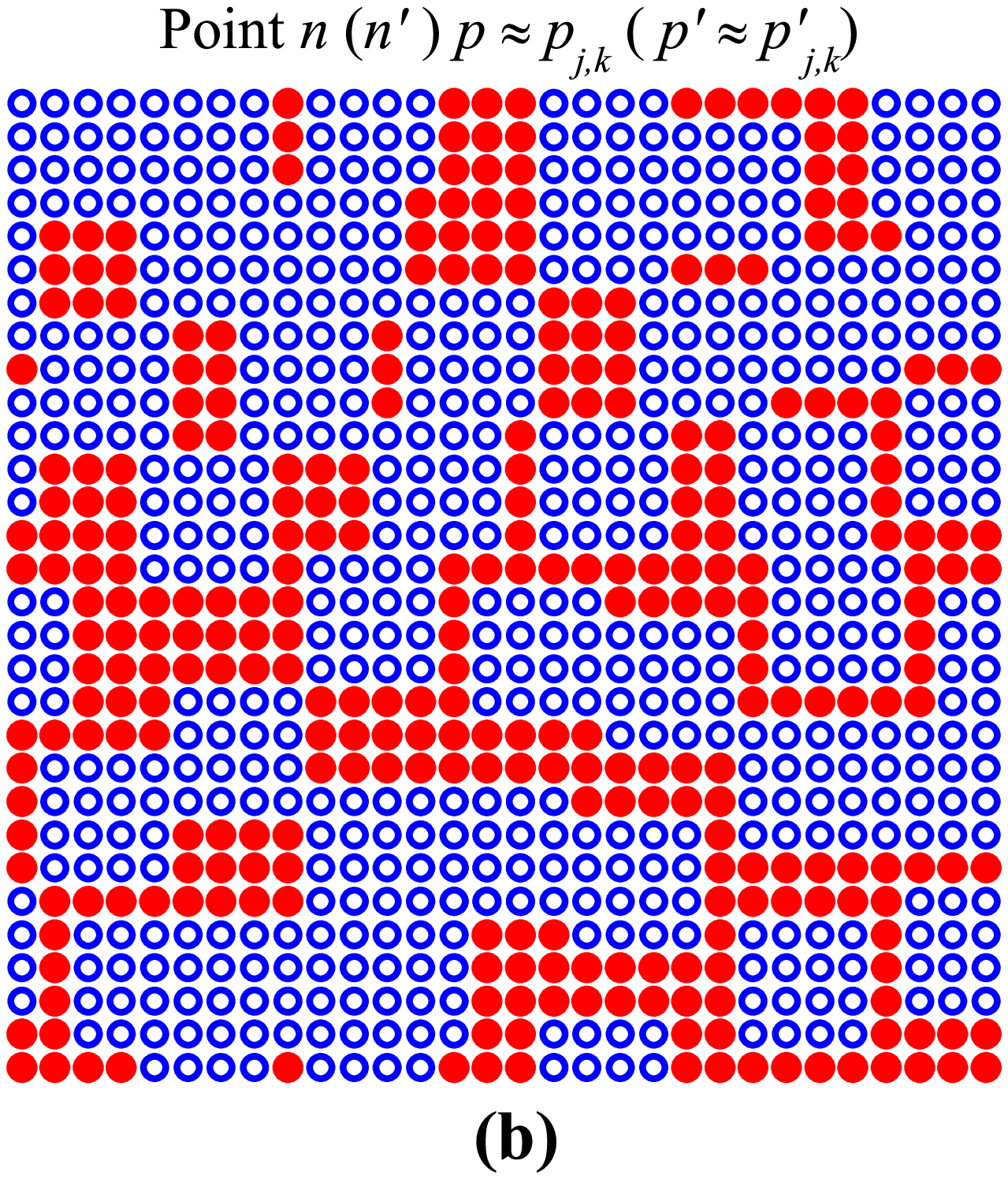}
\caption{(a) Snapshot of a typical lattice configuration in points $m$ and $m'$ of Fig. \ref{fig6}. (b) Same as part (a) for the points $n$ and $n'$ of Fig. \ref{fig6}. The meaning of red solid circles and blue open circles is given in the text. Thick line denotes a percolation path connecting two opposite sides of the lattice. 
\label{Figsnapk4}}
\end{center}
\end{figure}

The results for $p_{c,k}$ allows us to obtain the dependence of the inverse percolation threshold with $k$. The corresponding the curve is shown in Fig. \ref{fig6} (open squares). Figure \ref{fig6} also includes $p_{j,k}$ as a function of $k$ (Eq. (\ref{pjk}), solid squares).

$p_{c,k}$ shows a decreasing function with $k$ in the range $1 \leq k \leq 4$. The situation changes for $k \geq 5$ because all jammed configurations are percolating states, and consequently, there is no non-percolating phase in the whole range of allowed values of $p$. This finding means that the percolation phase transition disappears for $k \geq 5$. The last can be clearly seen in Fig. \ref{fig6}: for $k = 4$, the value for $p_{c,4}$ almost intersect the jamming curve. For bigger values of $k$, $p_{c,k}$ should be smaller than $p_{c,4}$ but those values are not possible to reach because of the blocking of the lattice. A similar behavior was observed for standard site percolation of $k^2$-mers on square lattices \cite{Nakamura86,Nakamura,PRE19}. In this case, the percolation phase transition disappears for $k \geq 4$. Above $k=3$, the jamming transition occurs before the system can reach the connectivity required for the formation of a percolating cluster.

In order to visualize better the differences (or asymmetry) between the classical percolation of $k^2$-mers on square lattices \cite{Nakamura86,Nakamura,PRE19} and the corresponding inverse percolation problem, the standard jamming and percolation curves have been included in Fig. \ref{fig6}. The values of jamming coverage ($p'_{j,k}$) and percolation threshold ($p'_{c,k}$) as a function of $k$ are shown as open and solid squares, respectively. The data correspond to results reported in Ref. \cite{PRE19}. While $p$ represent a fraction of occupied sites after removing $k^2$-mers from an initially fully occupied lattice, the nomenclature $p'$ indicates a fraction of occupied sites after depositing $k^2$-mers on an initially empty lattice. Thus, $p$($p'$) varies between 1(0) and $p_{j,k}$($p'_{j,k}$). 

In Fig. \ref{fig6}, five points have been marked as $i,j,l,m,n$ and five points have been marked as $i',j',l',m',n'$. The analysis of typical lattice configurations in these points will allow us to understand the observed differences between standard and inverse percolation problem. The corresponding snapshots are shown in Figs. \ref{Figsnapk3} (case $k=3$) and \ref{Figsnapk4} (case $k=4$). Each configuration can be thought of in two different ways. These two ways are: (1) the configuration was obtained by following a standard RSA process of $k^2$-mers. Then, blue open circles and red solid circles represent occupied sites by $k^2$-mers and empty sites, respectively; and (2) the configuration was obtained by removing $k^2$-mers as described in Section \ref{modelo}. Then, blue open circles and red solid circles represent empty sites (after removing $k^2$-mers) and occupied sites, respectively. Under these considerations, each snapshot in  
Figs. \ref{Figsnapk3} and \ref{Figsnapk4} allows us to discuss standard percolation [way (1)] and inverse percolation [way (2)]. Namely,

\begin{itemize}

\item Fig. \ref{Figsnapk3}(a), way (1), point $i'$ ($p' \approx 0.5$): The lattice is covered by small islands of occupied sites (blue open circles) and, accordingly, it does not exist a path of occupied sites connecting the opposite sides of the lattice. 

\item Fig. \ref{Figsnapk3}(a), way (2), point $i$ ($p \approx 0.5$): Even when half of the particles were removed from the lattice, a cluster of occupied sites  extends from one side to the opposite one of the system (see thick line). In other words, the percolating phase of occupied sites (red solid circles) has not been disconnected and, consequently, the inverse percolation phase transition has not happened yet.  

\item Fig. \ref{Figsnapk3}(b), way (1), point $j'$ ($p' \approx 0.61$): The fraction of occupied sites (blue open circles) is not sufficient to connect the opposite sides of the lattice (the standard percolation phase transition has not happened yet).

\item Fig. \ref{Figsnapk3}(b), way (2), point $j$ ($p \approx 0.39$): The fraction of occupied sites (red solid circles) is $p \approx 0.39 < p_{c,k=3}=0.40997(9)$. Then, the inverse percolation phase transition has happened, and it does not exist a path of occupied sites connecting the opposite sides of the lattice.

\item Fig. \ref{Figsnapk3}(c), way (1), point $l'$ ($p' \approx 0.65$): The fraction of occupied sites (blue open circles) is $p' \approx 0.659 > p'_{c,k=3}=0.63110(9)$ \cite{PRE19}. Then, the standard percolation phase transition has happened, and a path of occupied sites connects the opposite sides of the lattice (see thick line). 

\item Fig. \ref{Figsnapk3}(c), way (2), point $l$ ($p \approx 0.35$): As in the previous case [Fig.  \ref{Figsnapk3}(b)], $p \approx 0.35 < p_{c,k=3}=0.40997(9)$ and the percolating phase of occupied sites (red solid circles) has disappeared. 

\item Fig. \ref{Figsnapk4}(a), way (1), point $m'$ ($p' \approx 0.6$): The fraction of occupied sites (blue open circles) is not sufficient to percolate, accordingly, it does not exist a path of occupied sites connecting the opposite sides of the lattice. 

\item Fig. \ref{Figsnapk4}(a), way (2), point $m$ ($p \approx 0.4$): The fraction of occupied sites (red solid circles) is $p \approx 0.4 > p_{c,k=4}=0.36500(11)$. Then, a cluster of occupied sites extends from one side to the opposite one of the system (see thick line) and, consequently, the inverse percolation phase transition has not happened yet.

\item Fig. \ref{Figsnapk4}(a), way (1), point $n'$ ($p' \approx p'_{j,k}$): As expected from previous investigations \cite{Nakamura86,Nakamura,PRE19}, the jammed configuration shown in the figure is a nonpercolating state of occupied sites (blue open circles). In other words, the standard percolation phase transition disappears for $k^2$-mers on square lattices with $k \geq 4$. 

\item Fig. \ref{Figsnapk4}(a), way (2), point $n$ ($p \approx p_{j,k}$): The fraction of occupied sites (red solid circles) is $p \approx p_{j,k} < p_{c,k=4}=0.36500(11)$. Then, the inverse percolation phase transition has happened, and it does not exist a path of occupied sites connecting the opposite sides of the lattice. 

\end{itemize}

Summarizing, the detailed analysis presented in Figs. \ref{fig6}-\ref{Figsnapk4} indicates that (1) inverse and standard percolation phase transitions occur for $k=2$ and $k=3$; (2) only the inverse percolation phase transition occurs for $k=4$. The standard percolation phase transition disappears for $k \geq 4$; and (3) the inverse percolation phase transition is not possible for $k \geq 5$. 

It is also interesting to compare the results obtained for $k^2$-mers with the previously reported for straight rigid $k$-mers \cite{JSTAT2}. This comparison is shown in the inset of Fig. \ref{fig6}. Open squares (solid triangles) represent data obtained by removing $k^2$-mers ($k$-mers) from square lattices. Two important observations can be drawn from the figure: (1) while the phase transition disappears for $k^2$-mers with $k > 4$, percolating and non-percolating phases extend to infinity in the space of the parameter $k$ for rigid $k$-mers; and (2) the values of $p_{c,k}$ corresponding to $k^2$-mers remain below the curve obtained by removing $k$-mers.

In terms of network attacks, the behavior described in (1)-(2) indicates that the vulnerability of the network depends on the shape and size of the attacked region. Thus, extended attacks on linear sets of occupied sites are more effective than more compact attacks on $k \times k$ clusters of sites. As an illustrative example, it is necessary to remove almost $3/5$ of the sites to disconnect a network by removing sets of $2 \times 2$ square tiles of occupied sites. The same effect can be achieved by removing a little more than $2/5$ of the sites using dimers ($k$-mers with $k=2$). Moreover, for $k > 4$, the lattice remains connected even when the highest allowed concentration of removed $k^2$-mers is reached. The results shown in Fig. \ref{fig6} are consistent with those reported in Refs. \cite{Kornbluth,Lowinger}, where it was shown that the effectiveness of an attack depends on its degree of correlation.

\subsection{Critical exponents and universality}

In order to completely analyze the universality of the problem, the critical exponents $\beta$ and $\gamma$ can be obtained from the scaling behavior of the percolation order parameter $P=\langle S_L \rangle/L^2$ and the corresponding percolation susceptibility $\chi=\left( \langle S^2_L \rangle - \langle S_L \rangle^2 \right)/L^2$, respectively \cite{Stauffer}. $\langle \dots \rangle$ means an average over simulation runs. Thus,
\begin{equation}
P=L^{-\beta/\nu} \overline{P}\left( u \right), \label{functionP}
\end{equation}
where $u=| p - p_{c,k} | L^{1/\nu}$ and  $\overline{P}$ is the scaling function. At the point where $dP/dp$ is maximal, $u=$const. and
\begin{equation}
\left(\frac{dP}{dp}\right)_{\rm max}=L^{(-\beta/\nu+1/\nu)} \overline{P}\left( u \right) \propto L^{(1-\beta)/\nu}. \label{functionPmax}
\end{equation}

\begin{figure}
\includegraphics[scale=0.5]{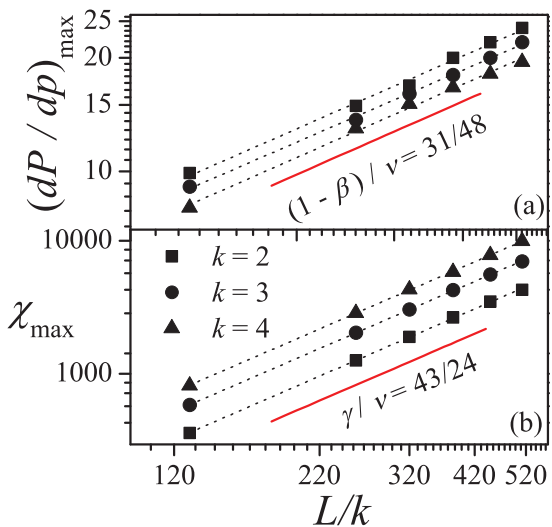}
\caption{ (a) Log-log plot of $\left(dP/dp\right)_{\rm max}$ as a function of $L/k$ for different values of $k$ as indicated. According to Eq. (\ref{functionP}), the slope of each curve corresponds to $(1-\beta)/\nu=31/48$. (b) Log-log plot of $\chi_{\rm max}$ as a function of $L/k$ for different values of $k$ as indicated. The slope of each line corresponds to $\gamma/ \nu=43/24$.
 \label{bx}}
\end{figure}

The exponent $\gamma$ can be determined by scaling the maximum value of the susceptibility. The behavior of $\chi$ at criticality is $\chi=L^{\gamma/\nu} \overline{\chi}\left( u' \right)$, where $u'=\left( p - p_{c,k} \right) L^{1/\nu}$ and $\overline{\chi}$ is the corresponding scaling function \cite{Stauffer}. At the point where $\chi$ is maximal, $u'=$const. and
$\chi_{\rm max} \propto L^{\gamma/\nu}$.

The data for $\left( dP/dp \right)_{\rm max}$ and $\chi_{\rm max}$ are shown in Figs. \ref{bx} (a) and \ref{bx} (b), respectively. The obtained values of the critical exponents are: $\beta=0.139(1)$ and $\gamma=2.38(2)$ ($k=2$); $\beta=0.141(5)$ and $\gamma=2.40(3)$ ($k=3$); and $\beta=0.138(1)$ and $\gamma=2.39(2)$ ($k=4$). So, as can be seen, simulation data are consistent with the exact values of the critical exponents of the ordinary percolation $\nu=4/3$, $\beta=5/36$ and $\gamma=43/18$, what clearly indicates that this problem belongs to the same universality class that the standard percolation problem \cite{PRE19}. This finding is expected since standard and inverse percolation problems are based on the RSA model, which has very short-range correlations.

\section{Conclusions}\label{conclu}

In this paper, the inverse percolation properties by removing square tiles composed by $k \times k$ occupied sites ($k^2$-mers) from square lattices has been studied by extensive numerical simulations complemented by finite-size scaling theory.

The inverse percolation problem deals with how connectivity varies during the dilution process of an initially fully occupied lattice. The procedure starts with an initial configuration in which  all sites of the lattice are occupied and, thereby, the opposite sides of the lattice are connected by nearest-neighbor occupied sites. Then, the system is diluted by  randomly removing $k^2$-mers from the surface following a RSA mechanism. In this framework, the jamming properties are studied while the main goal is to find the minimum concentration $p$ for which the connectivity disappears, this particular value is called inverse percolation threshold $p_{c,k}$.

On the other hand, $p_{j,k}$ is the coverage of the limit state, in which no more objects can be removed from the lattice due to the absence of clusters of nearest-neighbor with appropriate shape and size.  As it happens when multisite occupancy is considered, jamming coverage has a strong influence in the properties of the system. In this case, the jamming dependence on $k$ was calculated as: $p_{j,k}= 1-p'_{j,k}$ \cite{JSTAT2,JSTAT8}, where $p'_{j,k}$ is the jamming dependence for the standard RSA problem of $k^2$-mers on square lattices \cite{PRE19}.

In addition, the critical exponent characterizing the jamming process, $\nu_j$, was measured for different values of $k$. In all cases, the values obtained for $\nu_j$ remain close to 1, confirming that $\nu_j=2/d$ for RSA processes on $d$-dimensional Euclidean lattices. The scaling properties of 
the jamming probability were also investigated. By using data collapse analysis, we found that this quantity behaves at criticality as $W(p)=  \overline{W}\left[\left( p - p_{j,k} \right) L^{1/\nu_j}\right]$, where $\overline{W}$ is the corresponding scaling function. This scaling behavior of 
the jamming probability is reported here for the first time in the literature.

Once the limiting parameters $p_{j,k}$'s were determined, the percolation properties of the system were studied. We found that the percolation threshold has a monotonic decreasing dependence on $k$ and $p_{c,k}$ can only be obtained for $k=2$, $k=3$, and $k=4$. For  $k \geq 5$ all jammed configurations are percolating states and, consequently, the percolation transition disappears from $k\ge 5$. This implies that for larger values of $k$, the jamming critical concentration occurs before than the percolation phase transition and the system cannot be disconnected. This finding contrasts with the results previously obtained by removing straight rigid $k$-mers from square lattices \cite{JSTAT2}. In fact, in the case of linear $k$-mers, the percolation phase transition occurs for the whole range of $k$ sizes. Accordingly, percolating and non-percolating phases extend to infinity in the space of the parameter $k$.

The obtained results were also exhaustively compared with the ones corresponding to the standard jamming and percolation problem of $k^2$-mers on square lattices \cite{PRE19}. While the standard percolation phase transition disappears for $k \geq 4$, the inverse percolation phase transition still occurs for $k=4$. As mentioned in the paragraph above, the inverse percolation phase transition is not possible for $k \geq 5$. These findings indicate that, even though the jamming properties of the standard and inverse models are trivially symmetric, the inverse percolation problem can not be derived straightforwardly from the standard percolation problem and it deserves a detailed treatment as presented here.

It is interesting to analyze the results obtained for inverse percolation in terms of vulnerability and network attacks. In this context, the present study reinforces the concept that the vulnerability of the network depends on the shape and size of the attacked region. We found here that extended attacks on linear sets of occupied sites are more effective than more compact attacks on $k \times k$ clusters of sites. These results are consistent with those reported in Refs. \cite{Kornbluth,Lowinger}, where it was shown that the effectiveness of an attack depends on its degree of correlation.

Finally, the accurate determination of critical exponents ($\nu$, $\gamma$ and $\beta$) confirmed that the percolation phase transition involved in the system, which occurs for $k$ varying between 1 and 4, belongs to the same universality class as the standard percolation problem. 

Future efforts will be dedicated to developing a unifying work exploring the asymmetry between standard and inverse percolation, as well as global phase diagrams for different shapes and networks within the RSA model.

\section*{Acknowledgements}
This work was supported in part by CONICET (Argentina) under project number PIP 112-201101-00615; Universidad Nacional de San Luis (Argentina) under project No. 03-0816; and the National Agency of Scientific and Technological Promotion (Argentina) under project  PICT-2013-1678. The numerical work were done using the BACO parallel cluster (http://cluster\_infap.unsl.edu.ar/wordpress/) located  at Instituto de F\'{\i}sica Aplicada, Universidad Nacional de San Luis - CONICET, San Luis, Argentina.

\renewcommand{\baselinestretch}{1.0}

\end{document}